\pdfoutput=1

\documentclass[nofootinbib, aip, preprint, reprint]{revtex4-1}

\usepackage{amsmath, amsfonts, amsthm, amssymb, latexsym, mathrsfs, bm}

\usepackage[pdftex]{graphicx}
\usepackage[table]{xcolor}

\usepackage{tabularx}
\usepackage{multirow}

\newcolumntype{C}[1]{>{\centering\let\newline\\\arraybackslash\hspace{0pt}}m{#1}}
\newcolumntype{L}[1]{>{\raggedright\let\newline\\\arraybackslash\hspace{5pt}}m{#1}}

\renewcommand\figurename{Figure}
\renewcommand\tablename{Table}

\begin{document}
% change bibtex label
  \makeatletter
  \renewcommand\@biblabel[1]{#1. }
  \makeatother
% change figure font
  \makeatletter
  \renewcommand{\fnum@figure}[1]{\textbf{\figurename~\thefigure}.}
  \makeatother
% change table font
  \makeatletter
  \renewcommand{\fnum@table}[1]{\textbf{\tablename~\thetable}.}
  \makeatother
  
\title{
Path-accelerated molecular dynamics: Parallel-in-time integration using path integrals
}

\author{Jorge L. Rosa-Ra\'ices}
\affiliation{Division of Chemistry and Chemical Engineering, 
 California Institute of Technology, Pasadena, CA, 91125}

\author{Bin Zhang}
\affiliation{Department of Chemistry,
 Massachusetts Institute of Technology, Cambridge, MA, 02139}

\author{Thomas F. Miller III} 
\email[Correspondence: ]{tfm@caltech.edu}
\affiliation{Division of Chemistry and Chemical Engineering, 
 California Institute of Technology, Pasadena, CA, 91125}
 
\begin{abstract}
Massively parallel computer architectures create new opportunities for the performance of long-timescale molecular dynamics (MD) simulations.
Here, we introduce the path-accelerated molecular dynamics (PAMD) method that takes advantage of distributed computing to reduce the wall-clock time of MD simulation via parallelization with respect to MD timesteps. 
The marginal distribution for the time evolution of a system is expressed in terms of a path integral, enabling the use of path sampling techniques to numerically integrate MD trajectories.
By parallelizing the evaluation of the path action with respect to time  and by initializing the path configurations from a non-equilibrium distribution, the algorithm enables significant speedups in terms of the length of MD trajectories that can be integrated in a given amount of wall-clock time.
The method is demonstrated for Brownian dynamics, although it is generalizable to other stochastic equations of motion including open systems.
We apply the method to two simple systems, a harmonic oscillator and a Lennard-Jones liquid, and we show that in comparison to the conventional Euler integration scheme for Brownian dynamics, the new method can reduce the wall-clock time for integrating trajectories of a given length by more than three orders of magnitude in the former system and more than two in the latter.
This new method for parallelizing MD in the dimension of time can be trivially combined with algorithms for parallelizing the MD force evaluation to achieve further speedup.
\end{abstract}

\maketitle

\section{Introduction}

Molecular dynamics (MD)~\cite{Fre02, All17} is the central tool for simulating chemical, biological, and materials systems, with new algorithms and hardware expanding the range of accessible timescales and lengthscales~\cite{Dur11, Dro12, Vot02}.
Faster processors have played an important role in this expansion, although the most dramatic improvements in recent years have come from the number of available processors, rather than the clock-speed of the individual cores~\cite{Key07, Don11}.
In particular, highly multi-threaded computer architectures have been used to parallelize the MD force evaluation, greatly reducing the wall-clock time needed to perform an individual MD step~\cite{Pli95, Phi05, Sha09, Sal13, Pal15, Gro15}.
However, despite this progress in the parallelization of MD simulations with respect to the force evaluations (i.e., in space), less attention has been dedicated to the notion of parallelization with respect to the MD timesteps (i.e., in time).

The sequential nature of MD (i.e., the need to have access to a given timestep before the next timestep can be computed) would seem to discount the possibility of exploiting parallelization in time; nonetheless, methods for parallel-in-time integration are being developed and applied to MD simulation.
Most approaches~\cite{Lio01, Far03, Gar06, Emm12} are based on a prediction-correction paradigm that combines fine (i.e., accurate and expensive) and coarse (i.e., inaccurate and inexpensive) solvers to iteratively refine approximations of a trajectory in a convergent and parallel-in-time fashion.
A range of coarse solvers and iteration schemes have been employed to evaluate MD trajectories of molecular systems with parallelization in the time domain~\cite{Baf02, Yan06, Spe12, Byl13, Blu19}, leading to order-of-magnitude reductions in the wall-clock time-to-solution with respect to sequential integration at the fine level of accuracy.
Schemes for approximate long-timescale integration via trajectory splicing are an alternative route to parallelization in time, yielding accurate time evolution for systems that exhibit strong timescale separation on well-characterized regions of the potential energy landscape~\cite{Per16}.

The current work takes a different approach to parallelizing MD in time.
We demonstrate that by working with ensembles of trajectories in a path-integral framework, multiple processors can be employed to reduce the wall-clock time needed to evolve an MD trajectory of arbitrary length, without resorting to parallelization of the MD force evaluation.
This method of parallelization for MD trajectories is independent of, and thus entirely complementary to, parallelization of the MD force evaluations, and it creates new opportunities to harness large numbers of available computer processors for the generation of long-timescale MD trajectories.

\section{Method}

\subsection{MD integration based on path distributions}
\label{sec:theory}

In this work, we focus on the MD equation of motion governing Brownian (i.e., overdamped Langevin) dynamics
under potential $V$ at temperature $\beta^{-1}$,
\begin{equation}\label{eq:bd}
 \dot{x}(t) = -\gamma^{-1} \, V'(x(t)) + \sqrt{2D} \, \dot{w}(t),
\end{equation}
where the diffusion coefficient $D$ and the friction coefficient $\gamma$ are related by the Einstein relation $D=(\beta\gamma)^{-1}$, and $w(t)$ is the standard Wiener process.
MD trajectories can be generated by discretizing Eq.~\ref{eq:bd} with various numerical integration schemes~\cite{Bru84, Bra98, Ric03, Bus07, Bou14}, such as the forward Euler algorithm~\cite{All17}
\begin{equation}\label{eq:euler}
 x(t+dt) - x(t) = -{\gamma}^{-1} \, V'(x(t)) \, dt + \sqrt{2D dt} \, \xi,
\end{equation}
where $dt$ is the discretization timestep, and $\xi$ a standard Gaussian random variate.
The marginal distribution associated with time evolution of the system by $dt$ according to Eq.~\ref{eq:euler} is~\cite{Ris96}
\begin{align}\label{eq:marginaldt}
 K(&x(t+dt) | x(t); dt) \propto \nonumber \\
	&\exp \left\{ -\frac{dt}{4D} \left( \frac{x(t+dt)-x(t)}{dt} + \frac{V'(x(t))}{\gamma} \right)^2 \right\},
\end{align}
such that the likelihood of a MD trajectory of length $T = N\, dt$ that evolves the system along positions $\bm{X} = \{x(t_0), x(t_1), \ldots, x(t_N)\}$ at times $t_n = t + n\, dt$ is
\begin{equation}\label{eq:likelihood}
 \prod_{n=0}^{N-1} K(x(t_{n+1})|x(t_n); dt) \equiv e^{-S[\bm{X}]},
\end{equation}
where $S[\bm{X}]$ is the action associated with the MD trajectory.
From Eq.~\ref{eq:likelihood}, the position of the time-evolved system at time $T$ has a marginal distribution given by the path integral 
\begin{equation}\label{eq:marginalT}
 K(x(T+t)|x(t); T) \propto \int_\mathbb{R} \, \mathrm{d}x_1 \, \cdots \, \int_\mathbb{R} \mathrm{d}x_{N-1} \, e^{-S[\bm{X}]},
\end{equation}
where $x_n = x(t_n)$.
It is clear that this path-integral formulation of the ensemble of MD trajectories provides an equivalent description of the time evolution of the system as Eq.~\ref{eq:euler}.
Numerous studies have explored this path-integral formulation with variations of the underlying equation of motion and of the discretization of the action~\cite{Pra86, Ole96, Bol02, Mil07, Siv14}.

\begin{figure}[ht]
 \centering
 \includegraphics[width=0.45\textwidth]{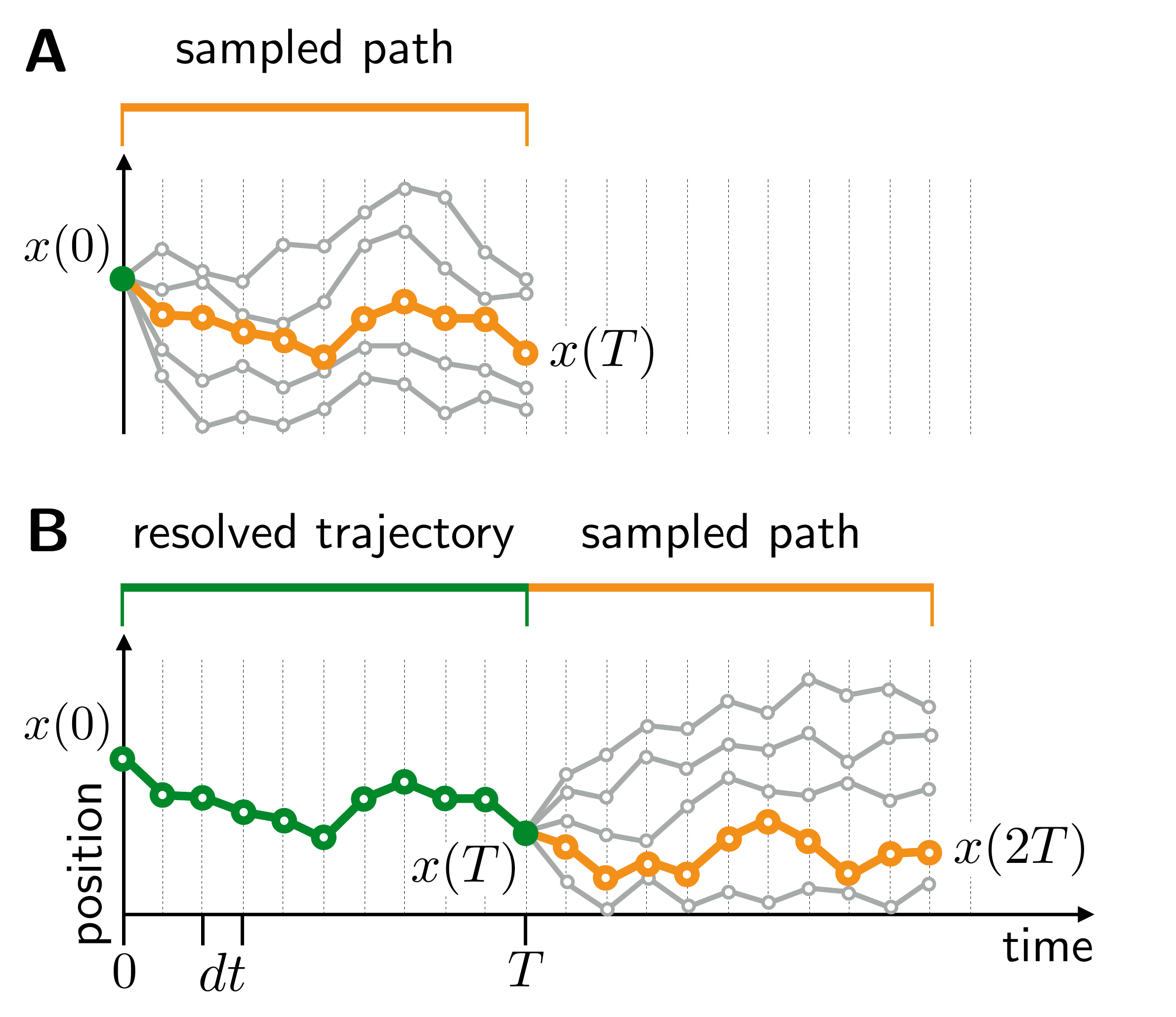}
 \caption{\label{fig:fig1}
  Illustration of a simple path-based MD integration scheme.
  \textsf{\textbf{A}} Sampling of the distribution of paths for the period of time from $0$ to $T$. A particular path drawn from this distribtion is indicated in orange.
  \textsf{\textbf{B}} Shifting along a sampled path (orange path in \textsf{\textbf{A}}) from $x(0)$ to $x(T)$, thereby resolving the segment of MD trajectory indicated in green, and then resuming the sampling of the paths for the period of time from $T$ to $2T$.
 }
\end{figure}

Setting aside issues of efficiency until section~\ref{sec:speedup}, we note that the path-integral formulation of the marginal distribution for the time-evolved system offers a simple MD integration scheme, illustrated in Fig.~\ref{fig:fig1}. First, \emph{sampling} from the distribution of paths of length $T$, with likelihood given by Eq.~\ref{eq:likelihood},
is performed using Monte Carlo (MC) or related methods (Fig.~\ref{fig:fig1}A)~\cite{Cep95, Stu04, Sto07}; by drawing a realization from this distribution, we obtain a segment of MD trajectory from time $0$ to time $T$ (illustrated by the heavy orange path in Fig.~\ref{fig:fig1}A).
Then, by \emph{shifting} from $x(0)$ to $x(T)$ along the sampled path, we resolve a trajectory from $x(0)$ to $x(T)$ (represented by the heavy green path in Fig.~\ref{fig:fig1}B) that is statistically equivalent to a realization from the Euler algorithm defined in Eq.~\ref{eq:euler}.
After shifting the tail of the path to $x(T)$, we restart the path sampling to extend the trajectory from time $T$ to time $2T$.
Iteration of this scheme will lead to the numerical integration of a MD trajectory of arbitrary length in time.

\begin{figure}[ht]
 \centering
 \includegraphics[width=0.45\textwidth]{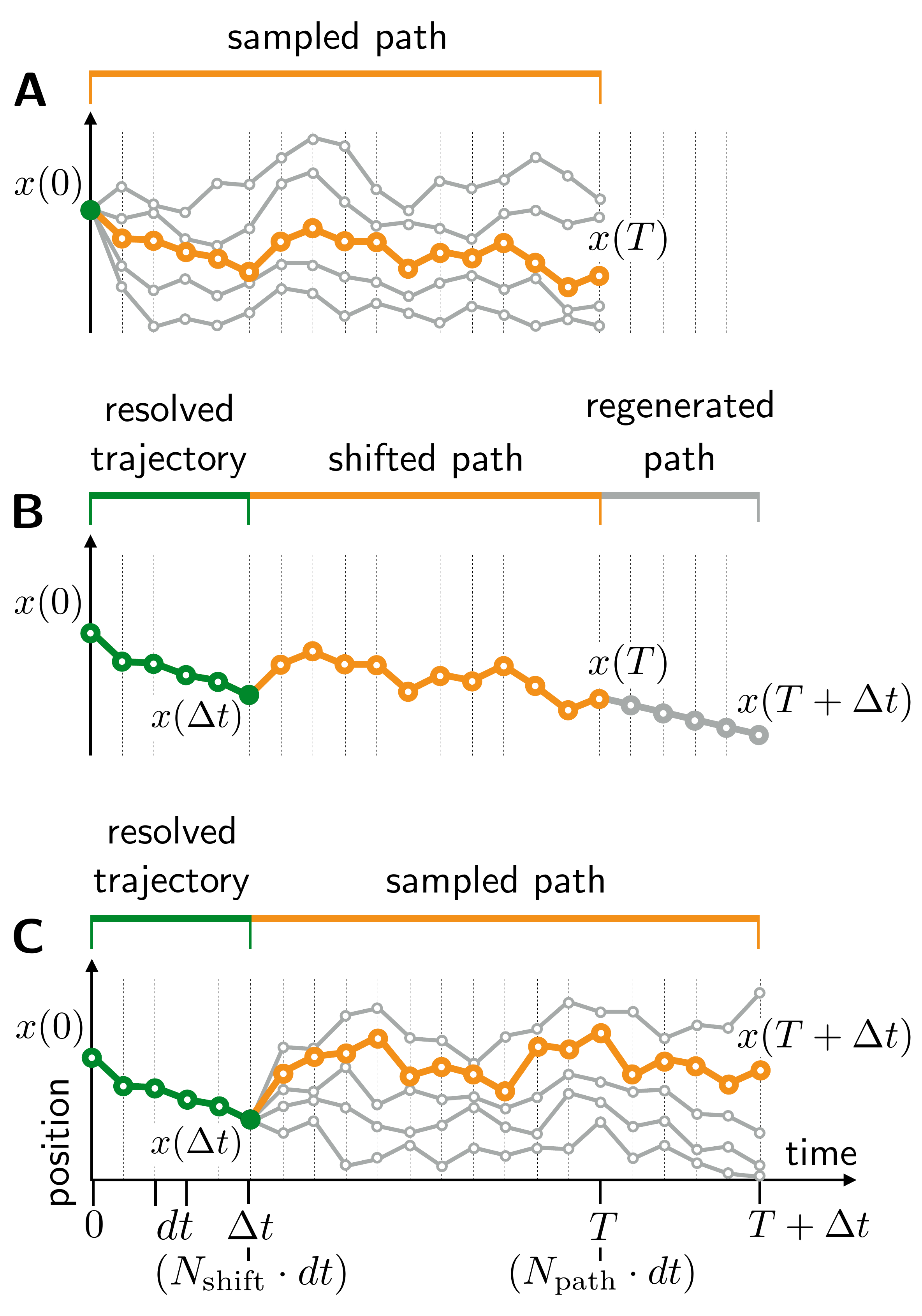}
 \caption{\label{fig:fig2}
  Illustration of a path-accelerated molecular dynamics (PAMD).
  \textsf{\textbf{A}} Sampling of the distribution of paths for the period of time from $0$ to $T$. A particular path drawn from this distribtion is indicated in orange.
  \textsf{\textbf{B}} Shifting along a sampled path (orange path in \textsf{\textbf{A}}) from $x(0)$ to $x(\Delta t)$, thereby resolving the segment of MD trajectory indicated in green, and regenerating the full length of the path by drawing positions for the system from time $T+dt$ to $T+\Delta t$ from an arbitrary distribution.
  \textsf{\textbf{C}} Sampling of the distribution of paths for the period of time from $\Delta t$ to $\Delta t+T$.
 }
\end{figure}

Fig.~\ref{fig:fig2} illustrates a generalized version of the integration scheme presented in Fig.~\ref{fig:fig1}.
Fig.~\ref{fig:fig2}A repeats Fig.~\ref{fig:fig1}A; we first sample a path of length $T$ that is discretized into $N_\mathrm{path}$ timesteps (where $N_\mathrm{path} = T / dt$) to obtain a realization of the path that is consistent with the marginal distribution of the time-evolved system for each time $\Delta t \le T$.
Then, in Fig.~\ref{fig:fig2}B, we shift the tail of the sampled path (indicated in orange) by $N_\mathrm{shift}$ timesteps (where $N_\mathrm{shift} = \Delta t / dt$) to the position $x(\Delta t)$.
With the remaining segment of the path now located at positions $\{x(\Delta t), x(\Delta t + dt), \ldots, x(T)\}$, we grow the path out of $x(T)$ by $N_\mathrm{shift}$ timesteps to \emph{regenerate} the original number of timesteps in the path.
The positions of the system at the regenerated timesteps can be drawn from any distribution (and in Fig.~\ref{fig:fig2}B they are obtained via straight-line extrapolation).
Finally, as illustrated in Fig.~\ref{fig:fig2}C, sampling is again performed to generate a path consistent with evolution from time $\Delta t$ to time $T + \Delta t$; this sampling removes any artifacts introduced by the arbitrary distribution used to grow the shifted path.
As for the scheme in Fig.~\ref{fig:fig1}, iteration of the scheme in Fig.~\ref{fig:fig2} yields an MD trajectory of arbitrary length in time that is statistically equivalent to a realization from the Euler algorithm.
The only difference between these two path-based integration schemes is that Fig.~\ref{fig:fig1} involves shifting along the full length of the sampled path, whereas Fig.~\ref{fig:fig2} involves shifting only a fraction of the way along the sampled path.

Just like the Euler scheme in Eq.~\ref{eq:euler}, the schemes illustrated in Figs.~\ref{fig:fig1} and \ref{fig:fig2} enable the numerical integration of MD trajectories. Each of these integration schemes consist of sequential iterations of an elementary step that predicts the state of the system at some later time.
In the Euler scheme, the prediction can be conducted analytically based on the distribution defined in Eq.~\ref{eq:marginaldt}.
In path-based integration schemes employing path lengths longer that $dt$, however, no such analytical expression exists for general systems; path sampling is therefore needed before each shifting event to generate time-evolved system positions consistent with the correct marginal distribution.

For the scheme in Fig.~\ref{fig:fig2}, it is assumed that the path distributions in parts A and C are well sampled.
For MC path sampling algorithms, this implies that the number of configurations of the path that are sampled in parts A and C, $N_\mathrm{sample}$, is large in comparison to the number that is needed to generate uncorrelated realizations of the path.
If $N_\mathrm{sample}$ is smaller than this decorrelation number, then the distribution of paths that is generated in part C may be biased by the way in which the path was regenerated in part B.
However, the only requirement for generating accurate MD trajectories using the scheme in Fig.~\ref{fig:fig2} is accurate sampling of paths consistent with the marginal distribution $K(x(\Delta t)|x(0); \Delta t)$; it is not essential that the marginal distribution associated with the full path, $K(x(T)|x(0); T)$, be sampled without error.
Recalling that $T = N_\mathrm{path}\, dt$ and $\Delta t = N_\mathrm{shift}\, dt$, this suggests that for a given path sampling algorithm, there is an interplay between parameters $N_\mathrm{sample}$, $N_\mathrm{path}$ and $N_\mathrm{shift}$; for given values of $N_\mathrm{path}$ and $N_\mathrm{shift}$, there is an associated number of path configurations ($N_\mathrm{sample}$) that must be sampled in order to generate a sufficiently accurate marginal distribution $K(x(\Delta t)|x(0); \Delta t)$.

\begin{figure*}[t]
 \centering
 \includegraphics[width=0.8\textwidth]{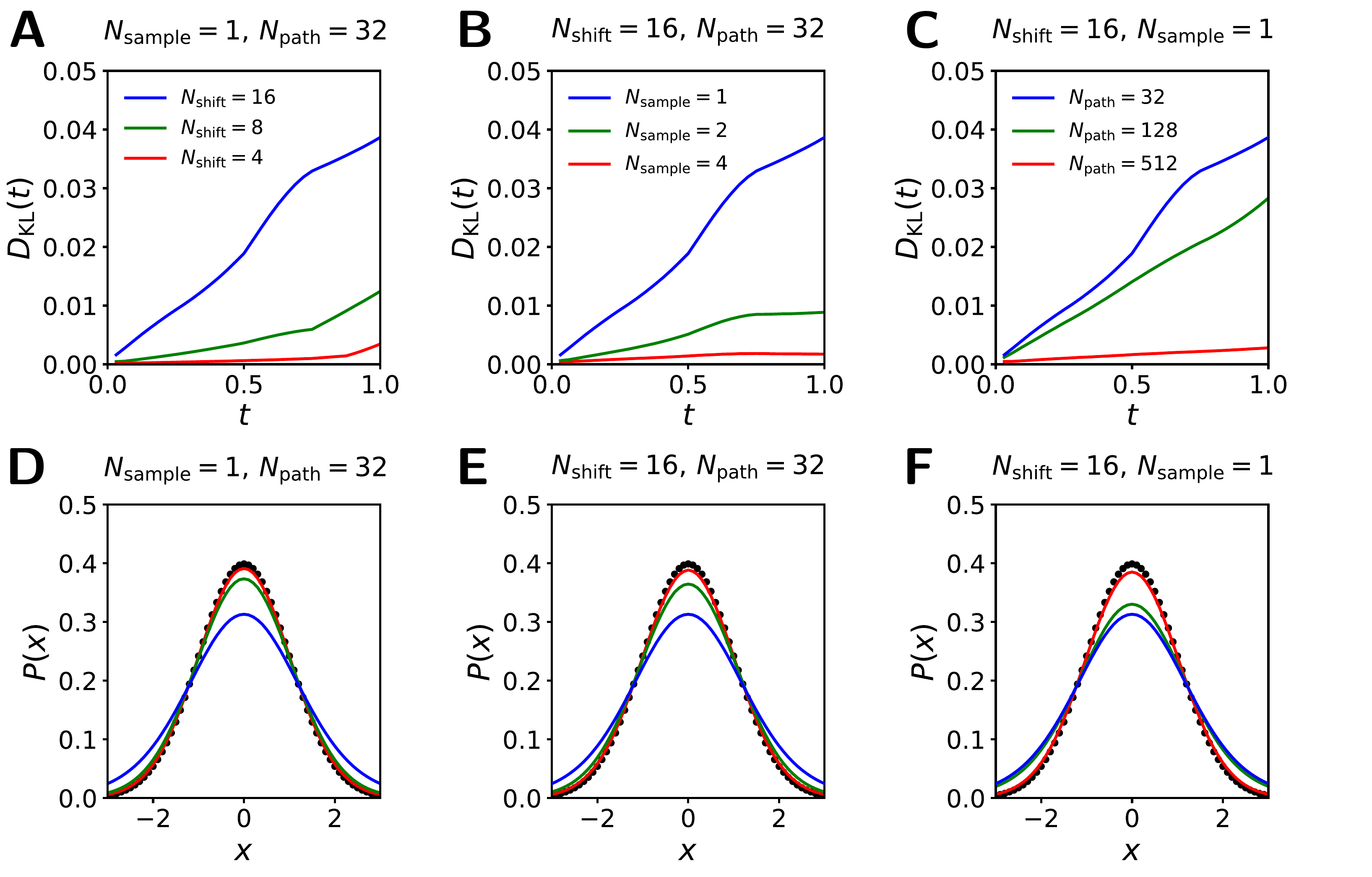}
 \caption{\label{fig:fig3}
  The interplay of the parameters $N_\mathrm{shift}$, $N_\mathrm{sample}$ and $N_\mathrm{path}$ in determining the accuracy of the PAMD integration scheme.
  Panels \textsf{\textbf{A}}, \textsf{\textbf{B}} and \textsf{\textbf{C}} show the Kullback-Leibler divergence $D_\mathrm{KL}(t)$ defined in Eq.~\ref{eq:dkl}, from the marginal distribution generated using the scheme in Fig.~\ref{fig:fig2} to the exact marginal distribution for the dynamics of an overdamped harmonic oscillator.
  Panels \textsf{\textbf{D}}, \textsf{\textbf{E}} and \textsf{\textbf{F}} plot the position distributions $P(x)$ sampled by the MD trajectories integrated using the numerically generated marginal distributions from panels \textsf{\textbf{A}}, \textsf{\textbf{B}} and \textsf{\textbf{C}}, respectively, in comparison to the exact position distribution shown in black dots.
  It is seen that the accuracy of the numerically generated marginal distribution dictates that of the integrated MD trajectory, and improves with decreasing $N_\mathrm{shift}$ (panels \textsf{\textbf{A}} and \textsf{\textbf{D}}), increasing $N_\mathrm{sample}$ (panels \textsf{\textbf{B}} and \textsf{\textbf{E}}) or increasing $N_\mathrm{path}$ (panels \textsf{\textbf{C}} and \textsf{\textbf{F}}) for given values of the remaining parameters (indicated at the top of each panel).
 }
\end{figure*}

This interplay between $N_\mathrm{sample}$, $N_\mathrm{path}$ and $N_\mathrm{shift}$ is illustrated in Figs.~\ref{fig:fig3}A-C, which plot the error in the marginal distribution generated using the scheme in Fig.~\ref{fig:fig2} for the Brownian dynamics of a harmonic oscillator.
Full calculation details are provided in Section~\ref{sec:calcdetails}.
The error plotted in Figs.~\ref{fig:fig3}A-C corresponds to the Kullback-Leibler divergence,
\begin{equation}\label{eq:dkl}
 D_\mathrm{KL}(t) = \left< \int_\mathbb{R} \mathrm{d}x_t \, P(x_t|x_0; t) \log \frac{P(x_t|x_0; t)}{Q(x_t|x_0; t)} \right>_{\!x_0},
\end{equation}
where $Q(x_t|x_0; t)$ is the marginal distribution estimated using sampled paths from the scheme in Fig.~\ref{fig:fig2}, and $P(x_t|x_0; t)$ is the exact marginal distribution.
The angled brackets denote averaging with respect to the Boltzmann distribution of %configurations
positions that is sampled by the exact dynamics, $P(x_0) = Z^{-1} e^{-\beta V(x_0)}$, where $Z = \int_\mathbb{R} \mathrm{d}x_0 \, e^{-\beta V(x_0)}$ is the partition function.
For a harmonic oscillator with potential $V(x) = \tfrac{1}{2} kx^2\;$~\cite{Gar09},
\begin{equation}\label{eq:marginalHO}
 P(x_t|x_0; t) \propto \exp \left\{ -\frac{\beta k}{2} \frac{(x_t - e^{-\gamma^{-1}kt}x_0)^2}{(1 - e^{-2\gamma^{-1}kt})} \right\},
\end{equation}
and we employ $k=1$ for the oscillator force constant, $\beta^{-1}=1$ for the temperature and $\gamma=1$ for the friction coefficient.
$D_\mathrm{KL}(t)$ returns non-negative values that approach $0$ as $Q(x_t|x_0; t)$ more accurately reproduces $P(x_t|x_0; t)$.
As a function of time $t$ along the sampled paths, $D_\mathrm{KL}(t)$ is plotted in Fig.~\ref{fig:fig3}A-C for seven simulations that employ the scheme in Fig.~\ref{fig:fig2} with different values of $N_\mathrm{sample}$, $N_\mathrm{path}$ and $N_\mathrm{shift}$.
The results correspond to sampled paths of length $T \ge 1$ that are discretized into timesteps of $dt=1/32$, and $D_\mathrm{KL}(t)$ is evaluated for the numerically generated marginal distributions at times $dt \le t \le 1$.

Fig.~\ref{fig:fig3}A addresses the case where $N_\mathrm{sample}$ and $N_\mathrm{path}$ are held fixed and various values of $N_\mathrm{shift}$ are used.
Comparison of the blue ($N_\mathrm{shift}=4$), green ($N_\mathrm{shift}=8$) and red ($N_\mathrm{shift}=16$) curves shows that for a given value of $N_\mathrm{sample}$, smaller values of $N_\mathrm{shift}$ lead to smaller errors in the numerically generated marginal distribution.
Using the scheme in Fig.~\ref{fig:fig2}, a given segment of the path is sampled $N_\mathrm{path} \cdot N_\mathrm{sample}/N_\mathrm{shift}$ times before it is used to generate the marginal distribution for the integration of the MD trajectory; therefore, smaller values of $N_\mathrm{shift}$ lead to better sampling of the path distribution and smaller errors in the marginal distribution.

Fig.~\ref{fig:fig3}B illustrates a second scenario where $N_\mathrm{shift}$ and $N_\mathrm{path}$ are held fixed and increasing values of $N_\mathrm{sample}$ are used.
Comparison of the blue ($N_\mathrm{sample}=1$), green ($N_\mathrm{sample}=2$) and red ($N_\mathrm{sample}=4$) curves shows that for a given value of $N_\mathrm{shift}$, larger values of $N_\mathrm{sample}$ (i.e., more sampling per shifting event) lead to smaller errors in the numerically generated marginal distribution for integrating the MD trajectory.
This result is intuitive, as more sampling leads to elimination of the bias associated with the arbitrary distribution used in the regeneration of the full length of the path.

In Fig.~\ref{fig:fig3}C, $N_\mathrm{shift}$ and $N_\mathrm{sample}$ are held fixed as the length of the sampled path ($N_\mathrm{path}$) is increased while keeping the path discretization timestep unchanged.
Comparison of the blue ($N_\mathrm{path}=32$), green ($N_\mathrm{path}=128$) and red ($N_\mathrm{path}=512$) curves demonstrates that increasing the total length $N_\mathrm{path} \cdot dt$ of the sampled paths improves the accuracy of the numerically generated marginal distribution.
Like decreasing $N_\mathrm{shift}$ for a given $N_\mathrm{path}$ (as in Fig.~\ref{fig:fig3}A), increasing $N_\mathrm{path}$ for a given $N_\mathrm{shift}$ allows for more sampling of each segment of the path employed to generate the marginal distribution associated with the MD time evolution.

While Figs.~\ref{fig:fig3}A-C illustrate the errors in the marginal distribution generated using the scheme in Fig.~\ref{fig:fig2}, Figs.~\ref{fig:fig3}D-F illustrate the corresponding errors in the equilibrium distribution that is sampled by the integrated MD trajectories.
For the various employed parameters, the results from path-based MD integration are compared to the exact Boltzmann distribution (dots) and, as expected, the errors in the marginal distribution with given values of $N_\mathrm{sample}$, $N_\mathrm{path}$ and $N_\mathrm{shift}$ are reflected in the distribution of positions that are visited in the MD trajectories.

In summary, Fig.~\ref{fig:fig3} demonstrates that decreasing $N_\mathrm{shift}$, increasing $N_\mathrm{sample}$, or increasing $N_\mathrm{path}$ leads to greater accuracy in the integrated MD trajectories; as will be shown in Section~\ref{sec:speedup}, the interplay between these three parameters is also critical for determining the computational efficiency of MD integration using the scheme in Fig.~\ref{fig:fig2}.

\begin{figure}[ht]
 \centering
 \includegraphics[width=0.45\textwidth]{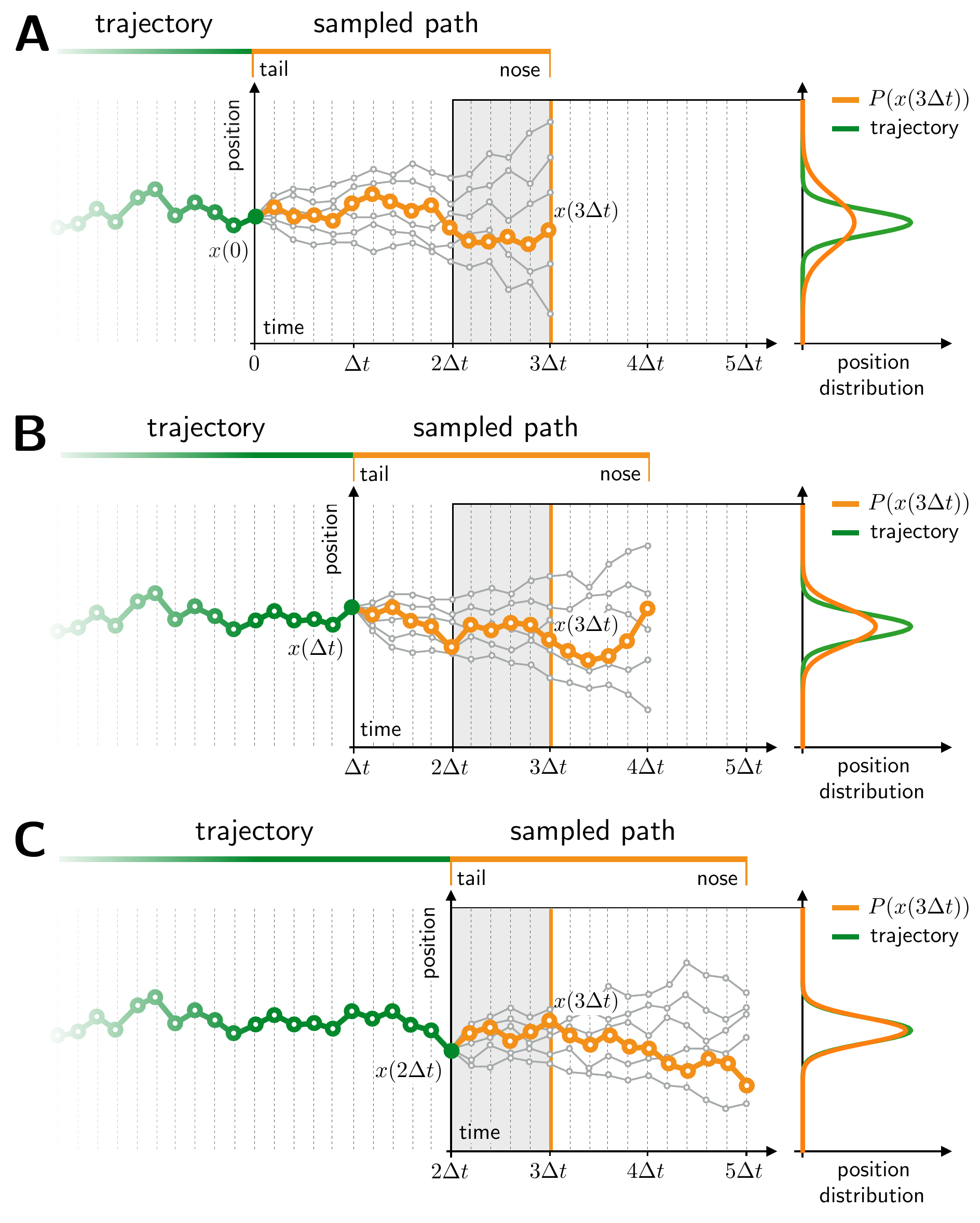}
 \caption{\label{fig:fig4}
  PAMD integrates equilibrium trajectories by relaxing non-equilibrium path segments.
  The state of the sampled path is shown for three consecutive iterations of the path-based integration scheme applied to a Brownian harmonic oscillator on the left of panels \textsf{\textbf{A}}, \textsf{\textbf{B}}, and \textsf{\textbf{C}}. The gray box in each panel highlights the configuration of a particular segment of the path after each iteration of the integration scheme.
  On the right of each panel are plotted the distributions $P(x(3\Delta t))$ of positions $x(3\Delta t)$ sampled by the boxed path segment at time $3\Delta t$ (orange curves), along with the distribution of configurations sampled by the integrated trajectory (green curve).
  The path segment inside the gray box in each panel reaches equilibrium by undergoing sampling as it shifts from the nose to the tail of the sampled path; accordingly, the distribution of sampled positions approaches that sampled by the integrated trajectory.
 }
\end{figure}

Before addressing efficiency, however, Fig.~\ref{fig:fig4} illustrates that the integration scheme in Fig.~\ref{fig:fig2} is a non-equilibrium relaxation process for the segments of the sampled path.
For the case of the harmonic oscillator, Figs.~\ref{fig:fig3}A-C indicate that errors in the numerically generated marginal distributions are typically larger at the nose of the sampled path.
This trend emerges because the integration scheme regenerates path segments in configurations that are out of equilibrium (Fig.~\ref{fig:fig2}B).
The light gray paths in Fig.~\ref{fig:fig4} correspond to independent realizations of the sampled path (orange) obtained while integrating a harmonic oscillator trajectory.
Since the segment at the nose of the path has undergone little sampling after regeneration, it is far from equilibrium with respect to the distribution of segments of an equilibrium harmonic oscillator trajectory (Fig.~\ref{fig:fig4}A; distributions at right).
However, as that segment works its way from the nose to the tail of the path, it is sampled with increasing accuracy (Figs.~\ref{fig:fig4}B and~C).
This relaxation process is illustrated by the distribution of positions, $P(x(3\Delta t))$, sampled by the foremost end $(x(3\Delta t))$ of the path segment in the grey box throughout Fig.~\ref{fig:fig4}; as the orange curves indicate, this distribution approaches that sampled by the harmonic oscillator trajectory (green curve) as the segment relaxes toward equilibrium and simultaneously approaches the tail of the sampled path.

\subsection{An opportunity for speedup}
\label{sec:speedup}

At face value, the path-based integration scheme in Fig.~\ref{fig:fig2} may appear to be inefficient, given the difficulties of sampling uncorrelated paths~\cite{Cep95, Del03, Mil07, Gin15}.
Yet, it has several potential advantages:
Firstly, there is an opportunity for parallelization, given that typical expressions for the path action incur a dominant source of computational cost from the evaluation of the forces in the system along the path ($V'(x)$ in Eq.~\ref{eq:marginaldt}). These forces can be evaluated independently, enabling straightforward parallelization of the action with respect to time.
Secondly, regeneration of the path to its full length following shifting (Fig.~\ref{fig:fig2}B) can be performed using an arbitrary distribution to obtain the system positions for the regenerated timesteps; consequently, it is possible to carry out this operation at a cost that is negligible relative to evaluation of the MD forces.
Thirdly, MC path sampling provides a numerically more stable way for generating trajectories than integration of the discretized equations of motion~\cite{Ole96}; thus, a sufficiently accurate MD trajectory may be obtained with the path-based integration scheme at a larger timestep than a conventional Brownian dynamics integrator would allow.

The above considerations suggest that the scheme in Fig.~\ref{fig:fig2} could lead to reduction of the wall-clock time associated with MD integration, in comparison with standard methods.
To quantify the speedup achieved with the new scheme, we introduce a measure $\chi$, with $\chi^{-1}$ defined as the number of force evaluations per processor per step of time $dt_\mathrm{E}$, where $dt_\mathrm{E}$ is the timestep used by the Euler algorithm to integrate the Brownian dynamics.
The wall-clock speedup of the path-based integration scheme is thus $\chi$, assuming that
\emph{(i)} evaluation of the MD forces dominates the cost of the evaluation of the path action,
\emph{(ii)} parallel computer processors are used to independently evaluate the forces along the discretized path, and
\emph{(iii)} regeneration of the full length of the path following shifting (Fig.~\ref{fig:fig2}B) is performed without evaluating the MD forces.
It is clear that for the Euler algorithm, $\chi = 1$, such that this measure provides a simple basis of comparison of the wall-clock time for the proposed path-based integration scheme (which employs parallelization in time) versus the wall-clock time for a conventional MD integration scheme (which does not).
In the current work, we set aside the complementary issue of speeding up MD integration via parallelization within the force evaluation.

For a general implementation of the integration scheme in Fig.~\ref{fig:fig2}, the expression for $\chi$ is obtained as follows.
Recalling previously introduced notation, we employ sampled paths of length $T$ that are discretized with a timestep of $dt$, which may be different (and is typically larger) than the numerically stable timestep for the Euler algorithm, $dt_\mathrm{E}$.
Let $N_\mathrm{force}$ be the number of MD force evaluations that are required during path sampling per shifting event, which depends on both $N_\mathrm{sample}$ and the details of the path sampling algorithm, and let $N_\mathrm{procs}$ be the number of employed parallel processors. 
Since the number of force evaluations per processor per shifting event is given by $N_\mathrm{force}/N_\mathrm{procs}$, the speedup is
\begin{equation}\label{eq:speedup}
 \chi = N_\mathrm{shift} \cdot \frac{N_\mathrm{procs}}{N_\mathrm{force}} \cdot \frac{dt}{dt_\mathrm{E}}.
\end{equation}
Eq.~\ref{eq:speedup} shows that the path-based integration scheme in Fig.~\ref{fig:fig2} offers the possibility for reduction of the wall-clock time needed to compute MD trajectories, relative to conventional MD.
Factors that enable this speedup include the increase in the discretization timestep ($dt$) relative to that possible for conventional MD ($dt_\mathrm{E}$), maximization of the number of integrated timesteps per shifting event ($N_\mathrm{shift}$), maximization of the number of parallel processors to perform the independent force evaluations associated with the calculation of the path action ($N_\mathrm{procs}$), and minimization of the number of force evaluations needed per shifting event ($N_\mathrm{force}$).
As will be shown in Section~\ref{sec:results}, this approach indeed enables substantial speedups in the integration of MD trajectories while preserving the accuracy of the dynamics, and we henceforth refer to the method as path-accelerated molecular dynamics (PAMD).

\section{Calculation Details}
\label{sec:calcdetails}

\begin{figure}[hb!]
 \centering
 \includegraphics[width=0.40\textwidth]{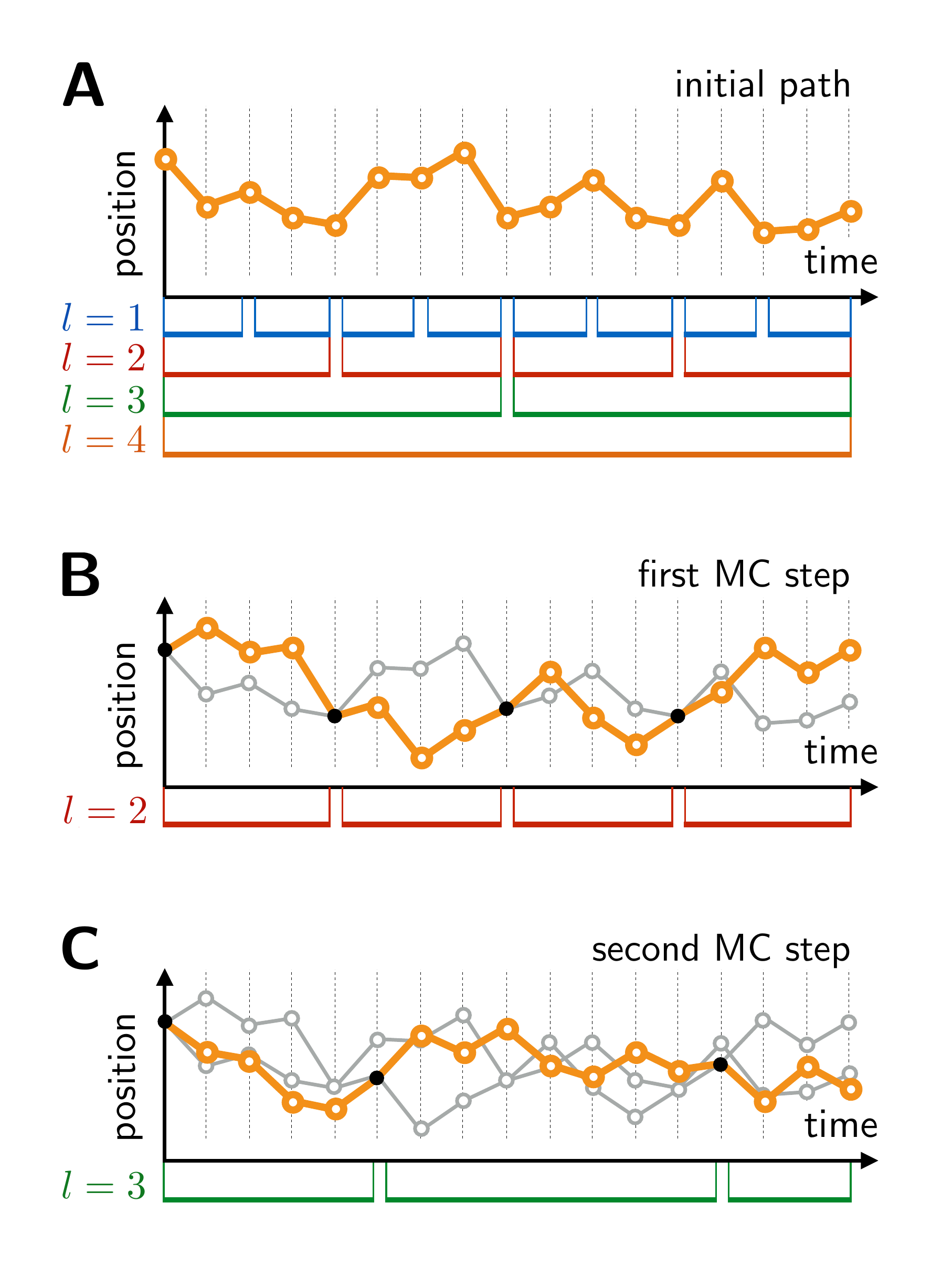}
 \caption{\label{fig:fig5}
  Illustration of the sliding and sampling algorithm.
  \textsf{\textbf{A}} Multilevel representation of a path with $N_\mathrm{path}=16$ timesteps ($L=4$ levels), shown for one Cartesian coordinate.
  An initial configuration of the path is shown in orange.
  \textsf{\textbf{B}} Update of the path configuration in \textsf{\textbf{A}} via a MC step at level $l=2$.
  Fixed system positions along the path are represented with black dots, configurations of the path before the update in gray, and the configuration of the path after the update in orange.
  \textsf{\textbf{C}} MC step at level $l=3$ following that shown in \textsf{\textbf{B}}.
  The new fragmentation of the path allows for updates of system positions that were held fixed in previous MC steps.
 }
\end{figure}

In the current work, we implement the PAMD method with sampling of the path distribution via the multilevel sliding and sampling algorithm for stochastic dynamics~\cite{Mil07}.
For a path of $N_\mathrm{path}$ timesteps, a total of $L = \log_2 N_\mathrm{path}$ levels are defined (Fig.~\ref{fig:fig5}A); finer levels (smaller values of the level index $1 \le l \le L$) correspond to partitions of the path into fragments of increasingly smaller length where the local configuration of the path is sampled.
In accordance with the sliding and sampling algorithm, neighboring path fragments share endpoints that are chosen randomly such that the length of the fragments varies from $1$ to $2^l$ timesteps; we call this random fragmentation.
For all fragmentations of the path at level $l$, internal fragments of the path are of length $2^l$ and fragments at the termini of the path have a combined length of $2^l$.
During a MC step for a given fragmentation of the path (Fig.~\ref{fig:fig5}B), the system positions at shared endpoints of neighboring path fragments are held fixed to permit mutually independent updates of the fragment configurations.
Furthermore, the position of the system at the tail endpoint of the path is always fixed throughout the MC step, whereas that at the nose endpoint of the path undergoes sampling together with the nose fragment.
Path fragment configurations are updated according to the Metropolis-Hastings criterion~\cite{Met53, Has70}, with trial configurations drawn from a distribution that satisfies the boundary conditions at the fragment endpoints.
Random fragmentation of the path is performed between MC steps, so that fixed system positions at previous fragment endpoints can be sampled during subsequent steps (Fig.~\ref{fig:fig5}C).

\begin{table*}[ht!]
 \centering
 \begin{minipage}{0.8\textwidth}
  \caption{\label{tbl:notation}
   Summary of notation employed to describe the path-based MD integration scheme introduced in the current work.
   The gray region corresponds to the definitions of parameters specific to the path sampling algorithm used.
  }
  \medskip
  \begin{tabular}{C{0.2\textwidth}|L{0.75\textwidth}}
   \hline
   $dt$                & timestep for discretization of the sampled path               \\
   $N_\mathrm{path}$   & number of timesteps in the sampled path                       \\
   $N_\mathrm{procs}$  & number of processors for parallel-in-time force evaluations   \\
   $N_\mathrm{force}$  & number of force evaluations per shifting event                \\
   $N_\mathrm{sample}$ & number of MC steps per shifting event                         \\
   $N_\mathrm{shift}$  & number of timesteps shifted                                   \\
   \hline
   \rowcolor{gray!25}
   $L$                 & total number of levels in the sampled path                    \\
   \rowcolor{gray!25}
   $l_\mathrm{min}$    & finest sampled level of the path                              \\
   \rowcolor{gray!25}
   $l_\mathrm{max}$    & coarsest sampled level of the path                            \\
   \hline
  \end{tabular}
 \end{minipage}
\end{table*}

At each MC step, a level is randomly selected between $l_\mathrm{min}$ and $l_\mathrm{max}$, with $1 \le l_\mathrm{min} \le l_\mathrm{max} \le L$.
The calculations reported here employ $l_\mathrm{min} > 1$ and $l_\mathrm{max} < L$, such that not all levels are directly sampled.
The choice of $l_\mathrm{max} < L$ corresponds to excluding the direct sampling of levels associated with long path fragments, on the basis of negligible acceptance.
The choice of $l_\mathrm{min} > 1$ corresponds to excluding the direct sampling of levels associated with short path fragments, as these are trivially updated via the direct sampling of longer fragments at coarser levels. 
Despite these choices, the sampling remains ergodic due to the random fragmentation of the path that occurs between MC steps.\cite{Mil07}

The distribution of paths used to generate trial configurations in each application is chosen to maximize the statistical efficiency of the sampling (i.e., minimize $N_\mathrm{sample}$) without requiring evaluation of the MD forces.
For the harmonic oscillator, trials are drawn from the distribution of free particle paths.
For simulations of the Lennard-Jones liquid, trials are drawn from the path distribution of a fluid of hard spheres with diameter $\sigma_\mathrm{HS}$; this strategy reduces the number of force evaluations needed to obtain likely Lennard-Jones path configurations by excluding those with high interparticle overlap from the ensemble of trial paths.
The likelihood of a path at the hard-sphere level is evaluated using an approximation of the pair propagator for diffusive hard spheres.\cite{Beh12}
For both applications, regeneration of the sampled path after shifting is performed with the same distribution used to generate trial configurations for the path sampling.

In total, each MC step involves a total number of $N_\mathrm{path}$ MD force evaluations.
Since $N_\mathrm{force}$ is defined as the number of force evaluations per shifting event, and since $N_\mathrm{sample}$ is the number of MC steps per shifting event, we have
\begin{equation}\label{eq:tfm1}
 N_\mathrm{force} = N_\mathrm{sample} \cdot N_\mathrm{path}.
\end{equation}
As we seek to maximize the wall-clock speedup via parallelization of these independent force evaluations, we employ one processor per force evaluation, and thus
\begin{equation}\label{eq:tfm2}
 N_\mathrm{procs} = N_\mathrm{path}.
\end{equation}
Additional parallelization within the force evaluation is of course possible, but is not considered in the current work.
Thus, we insert Eqs.~\ref{eq:tfm1} and \ref{eq:tfm2} into Eq.~\ref{eq:speedup} to arrive at the following expression for the PAMD speedup:
\begin{equation}\label{eq:speedupMC}
 \chi = \frac{N_\mathrm{shift}}{N_\mathrm{sample}} \cdot \frac{dt}{dt_\mathrm{E}}.
\end{equation}
See Table \ref{tbl:notation} for a summary of terms.

\section{Results}
\label{sec:results}

We now apply the PAMD algorithm to two model systems, considering time evolution according to Brownian dynamics (Eq.~\ref{eq:bd}) with $\gamma=1$ in appropriately reduced units.
Like the Euler algorithm (Eq.~\ref{eq:euler}), PAMD is a rigorous and formally exact way to integrate the dynamics of Eq.~\ref{eq:bd}, yet the numerical accuracy of the trajectories depends on the parameters employed.
In each application, we examine the relationship between the number of parallel processors employed and the speedup in the PAMD algorithm relative to the Euler algorithm ($\chi$ in Eq.~\ref{eq:speedupMC}), subject to the requirement that the MD trajectories integrated using both PAMD and the Euler algorithm preserve well-defined measures of accuracy.

In the current work, we focus exclusively on wall-clock speedups achieved via parallelization of the MD integration in time, setting aside the separate and complementary issue of parallelizing the force evaluation at each timestep. 
All reported speedups for PAMD in the current study are theoretical; they are obtained from Eq.~\ref{eq:speedupMC} under the stated assumptions.

\subsection{Harmonic oscillator}

Here, we consider the example of an overdamped harmonic oscillator, with potential $V(x) = \frac{1}{2}x^2$.
Two measures of the accuracy of the integrated MD trajectories are considered. The first reports on the degree to which the trajectories sample the correct equilibrium distribution,
\begin{equation}\label{eq:eeqHO}
 E_\mathrm{eq} =  \frac{\sqrt{\int_\mathbb{R} \mathrm{d}x \, \vert P(x) - P_\mathrm{s}(x) \vert^2}}{Z}
\end{equation}
where $P(x)$ is the exact Boltzmann distribution, $Z$ is the associated partition function, and $P_\mathrm{s}(x)$ is the equilibrium distribution of positions sampled by the numerical integration schemes.
The second measure of error reports on the accuracy of the MD time evolution. Specifically, we consider the autocovariance function
\begin{equation}\label{eq:tcf}
 C(t) = \left< x(t')x(t'+t) \right> = \lim_{T\rightarrow\infty}\frac{1}{T}\int_0^T \mathrm{d}t' \, x(t')x(t'+t),
\end{equation}
which is a simple exponential function for the overdamped harmonic oscillator~\cite{Gar09},
\begin{equation}\label{eq:tcfHO}
 C(t) = \left< x^2 \right> \exp(-\kappa t),
\end{equation}
where the angled brackets indicate Boltzmann averaging.
The second measure of error is thus
\begin{equation}\label{eq:edynHO}
 E_\mathrm{dyn} = \frac{\sqrt{(\kappa - \kappa_\mathrm{s})^2}}{k},
\end{equation}
where $\kappa=\gamma^{-1}=1$ is the exact decay constant and $\kappa_\mathrm{s}$ is the decay constant obtained by fitting the exponential decay of the autocovariance from the numerically integrated MD trajectories.
Specifically, $\kappa_\mathrm{s}$ is obtained by averaging over $100$ independent trajectories of length $10^5$ time units that are divided into $10^3$ non-overlapping time series, for which the log-autocovariance is linearly fit in the range $t \in [0, 4]$.
Simulation parameters for integration of the MD trajectories in this application are chosen to ensure that both measures of error remain below $3\%$.

\begin{figure}[hb]
 \centering
 \includegraphics[width=0.48\textwidth]{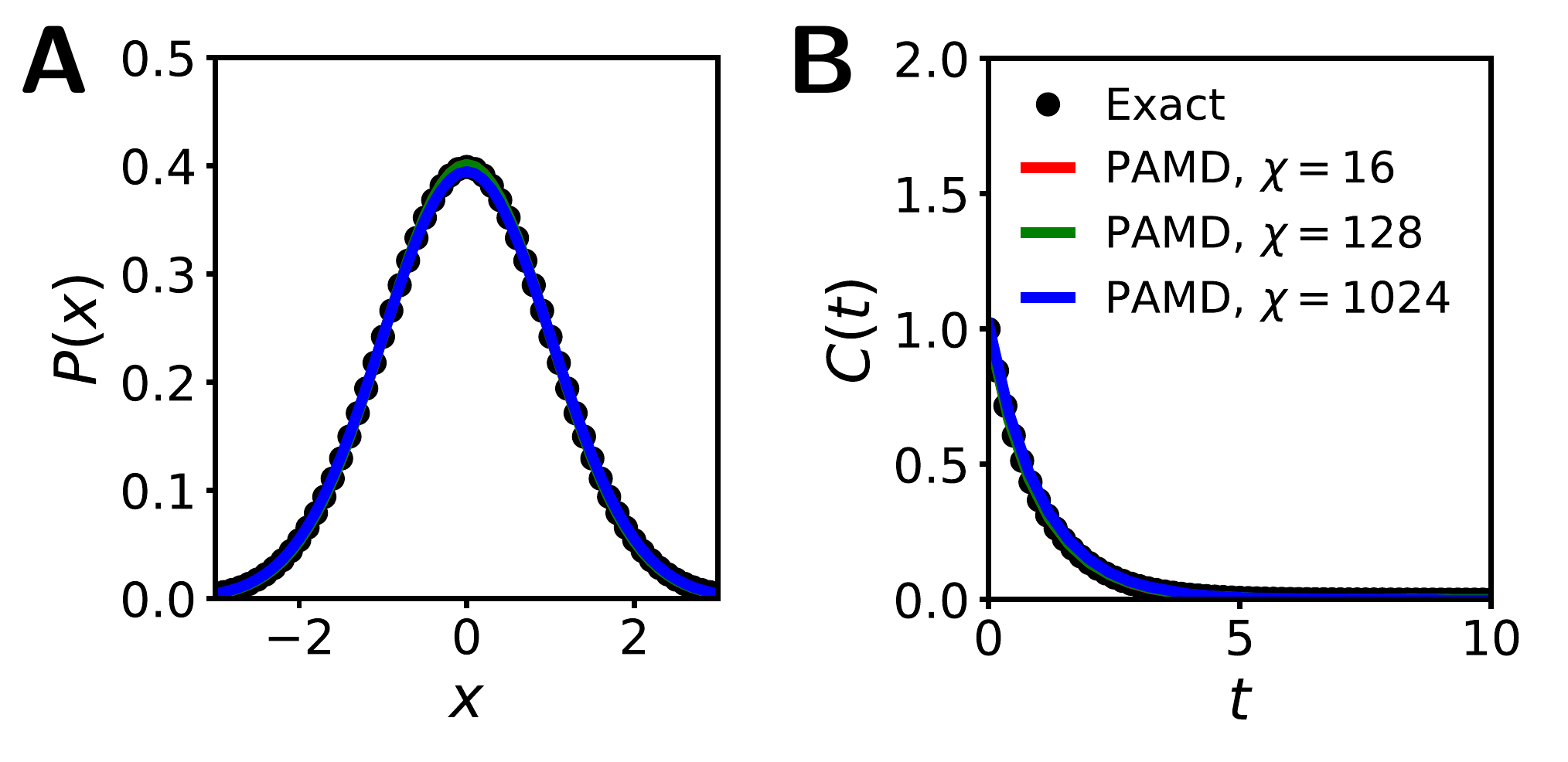}
\caption{\label{fig:fig6}
  For the harmonic oscillator, comparison of PAMD results (colored lines) with exact results (black dots) for
  \textsf{\textbf{A}} the Boltzmann distribution $P(x)$ and
  \textsf{\textbf{B}} the autocovariance function $C(t)$.
  The PAMD results correspond to Simulation $1$ (red), Simulation $2$ (green) and Simulation $3$ (blue) in Table~\ref{tbl:HOparam}, which respectively achieve speedups of $\chi=16$, $128$, and $1024$ relative to the Euler algorithm.
 }
\end{figure}

Table~\ref{tbl:HOparam} indicates that with a timestep of $dt_\mathrm{E}=0.025$, the Euler algorithm yields error values of $E_\mathrm{eq}=0.3\%$ and $E_\mathrm{dyn}=1.0\%$.
Also shown in the table are parameters for three separate PAMD simulations that obtain speedups of $\chi=16$, $128$ and $1024$ with respect to the Euler algorithm.
For both the equilibrium distribution and the autocovariance function, Fig.~\ref{fig:fig6} shows the comparison of the exact results and those obtained using PAMD with the aforementioned speedups.
The accuracy of the PAMD trajectories is clearly preserved in all simulations, as indicated by the plotted results and the reported values of $E_\mathrm{eq}$ and $E_\mathrm{dyn}$.

For all PAMD simulations reported in Table~\ref{tbl:HOparam}, a significant component of the speedup comes from the $16$-fold larger timestep that can be employed in the path-based scheme ($dt=0.4$ vs.\@ $dt_\mathrm{E}=0.025$).
The larger speedups achieved in Simulation $2$ ($\chi=128$) and Simulation $3$ ($\chi=1024$), in comparison to Simulation $1$ ($\chi=16$), arise from the larger ratios of $N_\mathrm{shift}$ to $N_\mathrm{sample}$ that are used in these simulations ($N_\mathrm{shift}/N_\mathrm{sample}=8$ for Simulation $2$ and $N_\mathrm{shift}/N_\mathrm{sample}=64$ for Simulation $3$) in comparison to Simulation $1$ (for which $N_\mathrm{shift}/N_\mathrm{sample}=1$).
The higher frequency of shifting events associated with larger values of the ratio $N_\mathrm{shift}/N_\mathrm{sample}$ places greater demand on the efficiency of the path sampling, and a larger number of timesteps in the sampled path ($N_\mathrm{path}$; hence, a larger number of parallel processors, $N_\mathrm{procs}$, per Eq.~\ref{eq:tfm2}) is needed to allow path segments to undergo a sufficient number of MC steps before they are used to generate the marginal distribution for the MD trajectories, as discussed in connection with Figs.~\ref{fig:fig3} and~\ref{fig:fig4}.

\begin{table}[ht!]
\centering
{
\begin{minipage}{0.45\textwidth}
\caption{\label{tbl:HOparam}
 Summary of PAMD simulation parameters used for the harmonic oscillator application.
 The gray region corresponds to the parameters specific to the path sampling algorithm used.
}
\medskip
\begin{tabular}{C{0.225\textwidth}|C{0.225\textwidth}C{0.225\textwidth}C{0.225\textwidth}}
\hline
& \multicolumn{3}{C{0.7\textwidth}}{Euler} \\
\hline
$dt_\mathrm{E}$  & \multicolumn{3}{C{0.7\textwidth}}{$0.025$} \\
$E_\mathrm{eq}$  & \multicolumn{3}{C{0.7\textwidth}}{$0.3\%$} \\
$E_\mathrm{dyn}$ & \multicolumn{3}{C{0.7\textwidth}}{$1.0\%$} \\ 
\hline
\hline
& \multicolumn{3}{C{0.7\textwidth}}{PAMD} \\
\hline
& Simulation $1$ & Simulation $2$ & Simulation $3$ \\
\hline
$dt$                  &    $0.4$ &    $0.4$ &    $0.4$ \\
$N_\mathrm{path}$     &     $16$ &    $256$ &   $4096$ \\
$N_\mathrm{sample}$   &      $2$ &      $1$ &      $1$ \\
$N_\mathrm{shift}$    &      $2$ &      $8$ &     $64$ \\
\hline
\rowcolor{gray!25}
$L$                   &      $4$ &      $8$ &     $12$ \\
\rowcolor{gray!25}
$l_\mathrm{min}$      &      $3$ &      $4$ &      $4$ \\
\rowcolor{gray!25}
$l_\mathrm{max}$      &      $3$ &      $5$ &      $8$ \\
\hline
$\chi$                &     $16$ &    $128$ &   $1024$ \\
$E_\mathrm{eq}$       &  $0.3\%$ &  $0.3\%$ &  $0.5\%$ \\
$E_\mathrm{dyn}$      &  $0.3\%$ &  $1.1\%$ &  $2.9\%$ \\
\hline
\end{tabular}
\end{minipage}
}
\end{table}

\subsection{Lennard-Jones liquid}

Here, we apply PAMD to a model for a molecular liquid. The pairwise interaction between particles is described using the standard cut-and-force-shifted Lennard-Jones potential,\cite{All17}
\begin{equation}\label{eq:ulj}
 U(r) = \left\{
 \begin{array}{rr}
  u(r) - u(r_\mathrm{c}) - (r - r_\mathrm{c}) \, u'(r_\mathrm{c}), & r \le r_\mathrm{c} \\
  0, & r > r_\mathrm{c}
 \end{array} \right., 
\end{equation}
where $u(r) = 4\epsilon \left\{ (\sigma/r)^{12} - (\sigma/r)^{6} \right\}$; throughout, we take $\epsilon=1$ and $\sigma=1$.
The system consists of $27$ particles placed in a cubic box at reduced density $\rho\sigma^{3} = 0.50$ and at constant reduced inverse temperature $\beta\epsilon = 0.74$.
Simulations are performed with periodic boundary conditions at constant volume, and the cutoff distance $r_\mathrm{c}$ corresponds to half of the simulation box-length.

\begin{figure}[ht!]
 \centering
 \includegraphics[width=0.45\textwidth]{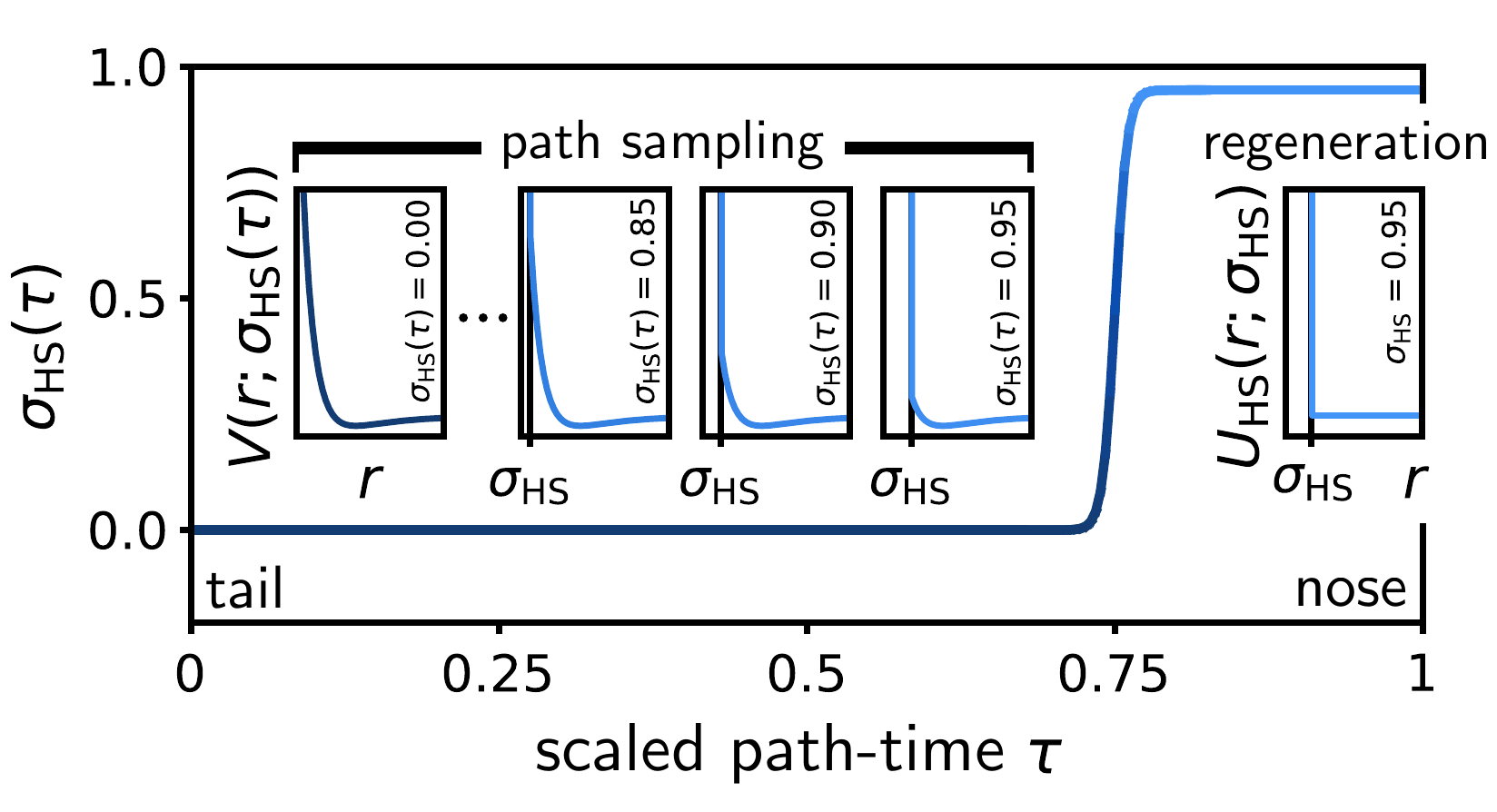}
 \caption{\label{fig:fig7}
  Hard-sphere schedule $\sigma_\mathrm{HS}(\tau)$, versus scaled path-time $\tau$, employed in the reported Lennard-Jones simulations.
  Insets show slices of the path-time dependent potential $V(r; \sigma_\mathrm{HS}(\tau))$ employed to sample path configurations (Eq.~\ref{eq:uljanneal}) at schedule values corresponding to biased ($\sigma_\mathrm{HS} \simeq 1$) and unbiased ($\sigma_\mathrm{HS} = 0$) sampling of the equilibrium path distribution, and the hard-sphere potential $U(r;\sigma_\mathrm{HS})$ used to regenerate path segments (Eq.~\ref{eq:uhs}).
 }
\end{figure}

As described in Section~\ref{sec:calcdetails}, the reported simulations for the Lennard-Jones liquid employ trial configurations drawn from a distribution of paths for a fluid of hard spheres with diameter $\sigma_\mathrm{HS}$.
To prevent the path-sampling bias from affecting the accuracy of the integrated trajectories, we employ a hard-sphere schedule that varies as a function of the path-time $\tau$, $\sigma_\mathrm{HS}(\tau)$; path configurations are then sampled in accordance with the path-time dependent potential
\begin{equation} \label{eq:uljanneal}
 V(r; \sigma_\mathrm{HS}(\tau)) = U(r) + U_\mathrm{HS}(r;\sigma_\mathrm{HS}(\tau)),
\end{equation}
where $U(r)$ is defined in Eq.~\ref{eq:ulj}, and $U_\mathrm{HS}(r; \sigma_\mathrm{HS})$ is the hard-sphere potential
\begin{equation} \label{eq:uhs}
 U_\mathrm{HS}(r; \sigma_\mathrm{HS}) = \left\{
 \begin{array}{rr}
  +\infty, & r \le \sigma_\mathrm{HS} \\
  0, & r > \sigma_\mathrm{HS}
 \end{array} \right..
\end{equation}
The schedule $\sigma_\mathrm{HS}(\tau)$ is chosen such that configurational volume is excluded at the nose of the path for enhanced sampling efficiency, and no volume is excluded at the tail of the path where the marginal distribution for MD integration is sampled (Fig.~\ref{fig:fig7}); in this way, path segments regenerated from the hard-sphere distribution are subsequently relaxed into the Lennard-Jones distribution as they shift from the nose to the tail of the path.

The accuracy of the integrated MD trajectories is evaluated in terms of the radial distribution function $g(r)$ and the self-diffusion coefficient $D$, using the respective error measures
\begin{equation} \label{eq:eeqLJ}
 E_\mathrm{eq} = \frac{\sqrt{\int_0^{r_\mathrm{c}} \mathrm{d}r\, \vert g(r) - g_\mathrm{s}(r) \vert^2}}{\int_0^{r_\mathrm{c}} \mathrm{d}r\, g(r)}
\end{equation}
and
\begin{equation} \label{eq:edynLJ}
 E_\mathrm{dyn} = \frac{\sqrt{(D - D_\mathrm{s})^2}}{D},
\end{equation}
where $g(r)$ and $D$ are reference quantities obtained using the Euler algorithm with a small timestep ($5 \times 10^{-5}$ Lennard-Jones time units), and $g_\mathrm{s}(r)$ and $D_\mathrm{s}$ are obtained using PAMD and the Euler algorithm with larger timesteps.
The diffusion coefficient is given by $D=\frac{1}{6} \lim_{t\rightarrow\infty} \partial_t \! \left<R^2(t)\right>$, where
\begin{equation} \label{eq:msdLJ}
 \left<{R}^2(t)\right> = \lim_{T\rightarrow\infty} \int_0^T \mathrm{d}t' \, \frac{1}{N} \sum_{i=1}^{N} \vert\mathbf{r}_i(t'+t)-\mathbf{r}_i(t')\vert^2
\end{equation}
is the mean-square displacement, $\mathbf{r}_i$ the position of the $i$th particle, and $N$ the number of particles~\cite{Fre02}.
$\left<R^2(t)\right>$ is obtained by averaging over $100$ independent trajectories that are divided into $100$ non-overlapping time series of length $1$, in Lennard-Jones time units, and a linear fit is performed in the range $t \in [0.2, 1]$ to evaluate $D_\mathrm{s}$ for the PAMD and Euler simulations.

\begin{figure}[ht]
 \centering
 \includegraphics[width=0.48\textwidth]{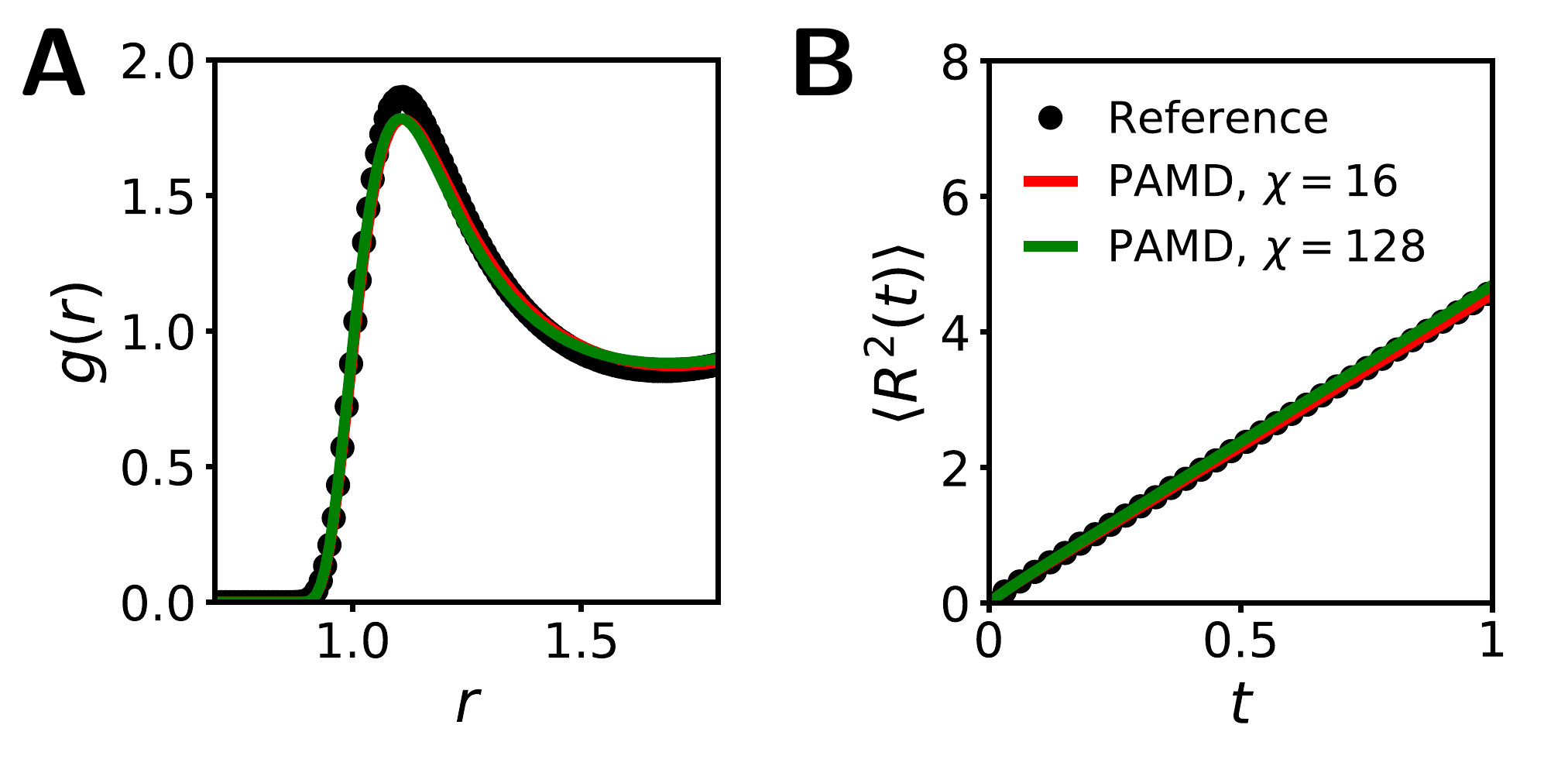}
 \caption{\label{fig:fig8}
  For the Lennard-Jones liquid at $\beta\epsilon = 0.74$ and $\rho\sigma^3 = 0.50$, comparison of PAMD results (colored lines) with numerically exact results (black dots) for
  \textsf{\textbf{A}} the radial distribution function and
  \textsf{\textbf{B}} the mean-square displacement.
  The PAMD results correspond to Simulation $1$ (red) and Simulation $2$ (green) in Table~\ref{tbl:LJparam}, which respectively achieve speedups of $\chi=16$ and $128$ relative to the Euler algorithm.
 }
\end{figure}

Table~\ref{tbl:LJparam} indicates that at a timestep of $dt_\mathrm{E} = 2.5 \times 10^{-4}$, the Euler algorithm yields error values of $E_\mathrm{eq}=1.4\%$ and $E_\mathrm{dyn}=0.1\%$; larger timesteps were found to lead to unstable Euler trajectories.
Also shown in Table~\ref{tbl:LJparam} are two PAMD simulations that lead to $16$-fold ($\chi=16$; Simulation $1$) and $128$-fold ($\chi=128$; Simulation $2$) reductions of the wall-clock time required to generate equivalently accurate MD trajectories for the Lennard-Jones liquid via the Euler algorithm, using simulation parameters that kept error values below $5\%$.
The radial distribution functions and mean-square displacements obtained from these two simulations are plotted with the corresponding reference quantities in Figs.~\ref{fig:fig8}A and~B.
Excellent agreement between the PAMD and reference quantities is evident in the plots and from the values of $E_\mathrm{eq}$ and $E_\mathrm{dyn}$ reported in Table~\ref{tbl:LJparam}.

As in the harmonic oscillator application, the speedups reported in Table~\ref{tbl:LJparam} for the Lennard-Jones liquid are partially enabled by the use of a larger timestep in PAMD ($dt = 5 \times 10^{-4}$) than is possible for stable numerical integration via the Euler method ($dt_\mathrm{E} = 2.5 \times 10^{-4}$).
The remaining speedup in both simulations comes from using shift lengths that integrate $N_\mathrm{shift}=8$ (Simulation $1$) and $N_\mathrm{shift}=64$ (Simulation $2$) timesteps of MD trajectory at a rate of $N_\mathrm{sample}=1$ path-sampling
steps per shifting event.
Accurate integration at these speedups requires efficient sampling of path modes that are commensurate with the shifting timescale ($8\, dt$ in Simulation $1$ and $64\, dt$ in Simulation $2$); accordingly, long paths ($N_\mathrm{path}=128$ in Simulation $1$ and $N_\mathrm{path}=4096$ in Simulation $2$) are employed in both simulations.

\begin{table}[ht]
\centering
{
\begin{minipage}{0.45\textwidth}
\caption{\label{tbl:LJparam}
 Summary of PAMD simulation parameters used for the application to the Lennard-Jones liquid.
 The gray region corresponds to the parameters specific to the path sampling algorithm used.
}
\medskip
\begin{tabular}{C{0.25\textwidth}|C{0.25\textwidth}C{0.25\textwidth}}
\hline
& \multicolumn{2}{C{0.5\textwidth}} {Euler} \\
\hline
$dt_\mathrm{E}$  & \multicolumn{2}{C{0.5\textwidth}}{$2.5 \times 10^{-4}$}  \\
$E_\mathrm{eq}$  & \multicolumn{2}{C{0.5\textwidth}}{$1.4\%$}  \\
$E_\mathrm{dyn}$ & \multicolumn{2}{C{0.5\textwidth}}{$0.1\%$}  \\ 
\hline
\hline
& \multicolumn{2}{C{0.5\textwidth}} {PAMD} \\
\hline
& Simulation $1$ & Simulation $2$ \\
\hline
$dt$                & $5 \times 10^{-4}$ & $5 \times 10^{-4}$ \\
$N_\mathrm{path}$   &    $128$ &    $4096$ \\
$N_\mathrm{sample}$ &      $1$ &       $1$ \\
$N_\mathrm{shift}$  &      $8$ &      $64$ \\
\hline
\rowcolor{gray!25}
$L$                 &      $7$ &      $12$ \\
\rowcolor{gray!25}
$l_\mathrm{min}$    &      $3$ &       $4$ \\
\rowcolor{gray!25}
$l_\mathrm{max}$    &      $5$ &       $7$ \\
\hline
$\chi$              &     $16$ &     $128$ \\ 
$E_\mathrm{eq}$     &  $3.5\%$ &   $3.7\%$ \\
$E_\mathrm{dyn}$    &  $1.0\%$ &   $1.9\%$ \\
\hline
\end{tabular}
\end{minipage}
}
\end{table}

\section{Conclusions}

The field of MD simulation faces important challenges in harnessing massively parallel computer architectures.
Although successful parallelization of the the force evaluation can be expected as the system size grows (i.e., weak scaling), there exists a much more difficult challenge of employing ever-larger numbers of parallel processors to accelerate the simulation of systems of a fixed size (i.e., strong scaling).
Remarkable success has been achieved in this vein~\cite{Pli95, Phi05, Sha09, Sal13, Pal15, Gro15}, but fundamental limitations are inevitable.

The current work suggests that parallelization in the dimension of time via path integrals offers a promising avenue for future progress.
We introduce the PAMD approach, which enables significant speedups over conventional Brownian dynamics algorithms via parallelization of the path-action with respect to time.
Proof-of-principle applications demonstrate that the algorithm can be applied straightforwardly to the harmonic oscillator and the Lennard-Jones liquid, where speedups of up to three orders of magnitude over the conventional Euler integration scheme for Brownian dynamics are achieved.
For a large class of systems including the two examples studied here, we suspect that even greater speedups are possible with the use of larger numbers of parallel processors and enhancement of the MC path sampling efficiency.

Although promising for the systems presented here, the PAMD approach will likely require additional methodological developments to become applicable to long-timescale, large-scale simulations.
Central to this effort will be the refinement of path sampling methodologies that lead to the reduction in the number of parallel processors that are needed for a given amount of speedup with the method.
Regardless, we feel that the natural parallelization of path-integral formulations, combined with the increasing availability of massively parallel computer resources, should motivate increased attention to the opportunities of parallelizing molecular dynamics simulation in time.

\begin{acknowledgments}
 We gratefully acknowledge stimulating discussions with Matthew G.\@ Welborn, Eric Vanden-Eijnden and Gavin E.\@ Crooks.
 This work was supported in part by the Department of Energy under Award No. DE-FOA-0001912 and the Office of Naval Research under Award No. N00014-10-1-0884.
\end{acknowledgments}

\bibliography{pamd}

\clearpage
\newpage

\end{document}